\documentclass[aps,prl,twocolumn,superscriptaddress]{revtex4-2}

\usepackage{graphicx}
\usepackage{latexsym}
\usepackage{amsmath,amssymb}
\usepackage{xcolor}
\usepackage[utf8]{inputenc}
\usepackage[english]{babel}

\graphicspath{{figures_main/}}

\usepackage{xr-hyper}
\usepackage{hyperref}

\hypersetup{
    colorlinks=true,
    linkcolor=blue,
    filecolor=magenta,      
    urlcolor=red,
    pdftitle={Manuscript MoS2Au},
    pdfpagemode=FullScreen,
    }

\usepackage{romannum}

\makeatletter
\newcommand*{\addFileDependency}[1]{
  \typeout{(#1)}
  \@addtofilelist{#1}
  \IfFileExists{#1}{}{\typeout{No file #1.}}
}
\makeatother

\newcommand*{\myexternaldocument}[1]{%
    \externaldocument{#1}%
    \addFileDependency{#1.tex}%
    \addFileDependency{#1.aux}%
}

\myexternaldocument{suppl}

\begin{document}
\title{Twist angle dependent electronic properties of exfoliated single layer MoS$_2$ on Au(111)}

\author{Ishita Pushkarna}
\affiliation{DQMP, Université de Genève, 24 Quai Ernest Ansermet, CH-1211 Geneva, Switzerland}
\affiliation{These authors contributed equally to this work.}
\author{\'{A}rp\'{a}d P\'{a}sztor}
\email{arpad.pasztor@unige.ch}
\affiliation{DQMP, Université de Genève, 24 Quai Ernest Ansermet, CH-1211 Geneva, Switzerland}
\affiliation{These authors contributed equally to this work.}
\author{Christoph Renner}
\email{christoph.renner@unige.ch}
\affiliation{DQMP, Université de Genève, 24 Quai Ernest Ansermet, CH-1211 Geneva, Switzerland}

\date{\today}

\begin{abstract}
Synthetic materials and heterostructures obtained by the controlled stacking of exfoliated monolayers are emerging as attractive functional materials owing to their highly tunable properties. We present a detailed scanning tunneling microscopy and spectroscopy study of single layer MoS$_2$-on-gold heterostructures as a function of twist angle. We find that their electronic properties are determined by the hybridization of the constituent layers and are modulated at the moiré period. The hybridization depends on the layer alignment and the modulation amplitude vanishes with increasing twist angle. We explain our observations in terms of a hybridization between the nearest sulfur and gold atoms, which becomes spatially more homogeneous and weaker as the moiré periodicity decreases with increasing twist angle, unveiling the possibility of tunable hybridization of electronic states via twist angle engineering. 
\end{abstract}


\maketitle

The rapid increase in the number of 2D materials which can be exfoliated and the fantastic progress in their layer-by-layer stacking with controlled sequence \cite{Kim2017} and twist angle \cite{Naimer2021, Pei2022} open up exceptional opportunities to design new functional quantum materials. A famous example is twisted bilayer graphene, whose very rich phase diagram ranges from correlated insulating to superconducting phases depending on twist angle and electrostatic doping \cite{Cao2018,Sanchez2012}. Heterostructures with unique properties can also be obtained by stacking 2D exfoliated transition metal dichalcogenides (TMD) \cite{Geim2013, Chen2020, Lin2014, Rhodes2019}.

Exfoliated 2D materials in direct proximity to a metal surface with selected twist angles offer further attractive materials engineering perspectives  \cite{Wu2022}. Such structures are challenging to prepare with a clean surface suitable for scanning tunneling microscopy (STM). Therefore, many STM studies to date have been carried out on epitaxial thin films grown in-situ by molecular beam epitaxy (MBE) or by chemical vapor deposition (CVD). The twist angle with the substrate of such films is set by thermodynamics and cannot be tuned at will. It is nearly 0$^\circ$ for 2H-MoS$_2$ evaporated on Au(111) \cite{Gronborg2015, Yasuda2017, Zhou2016ACSNano}, one of the most studied TMD on a metallic substrate \cite{Yang2020, Li2021}. 

Manual stacking of exfoliated monolayers allows to select arbitrary twist angles between the 2D material and the metallic substrate. Only a few STM studies of exfoliated 2H-MoS$_2$ monolayers (hereafter simply MoS$_2$) on Au(111) have been published \cite{Wu2020, Blue2020, Peto2019, Vancso2016, Peto2018, Magda2015, Hus2021}, without any strong focus on twist angle dependent properties. Here, we present a detailed STM and scanning tunneling spectroscopy (STS) investigation of the electronic properties of MoS$_2$ on Au(111) as a function of twist angle. We find that the semiconducting gap and band edges are modulated at the moir\'e period, with a modulation amplitude that vanishes with increasing twist angle. We explain our observations in terms of a hybridization between the sulfur and gold atoms, which becomes spatially more homogeneous and weaker as the moiré periodicity decreases with increasing twist angle.

\section{RESULTS}
We performed STM and STS on continuous, millimeter-sized monolayer (ML) flakes obtained by exfoliating 2H-MoS$_2$ \cite{Huang2020} onto template-stripped Au substrates \cite{Hegner1993}. These substrates are poly-crystalline Au(111) films stripped from an ultra-flat silicon wafer (see Methods). STM and X-ray diffraction measurements show that they consist of Au(111) orientated grains, where different grains can be slightly tilted and rotated about their [111]-axis (see SI section~\Romannum{1}). ML MoS$_2$ on Au(111) is identified by its characteristic Raman spectrum \cite{Yasuda2017, Velicky2020, Pollmann2021} and from optical images where MoS$_2$ appears darker on a bright Au substrate \cite{velicky2018} (see SI section~\Romannum{1}). A further confirmation that we have ML MoS$_2$ is the observation of a moir\'e pattern in STM topography, which is absent for bilayer and thicker MoS$_2$ flakes \cite{Gronborg2015}. 

\begin{figure*}[htp]

    \includegraphics{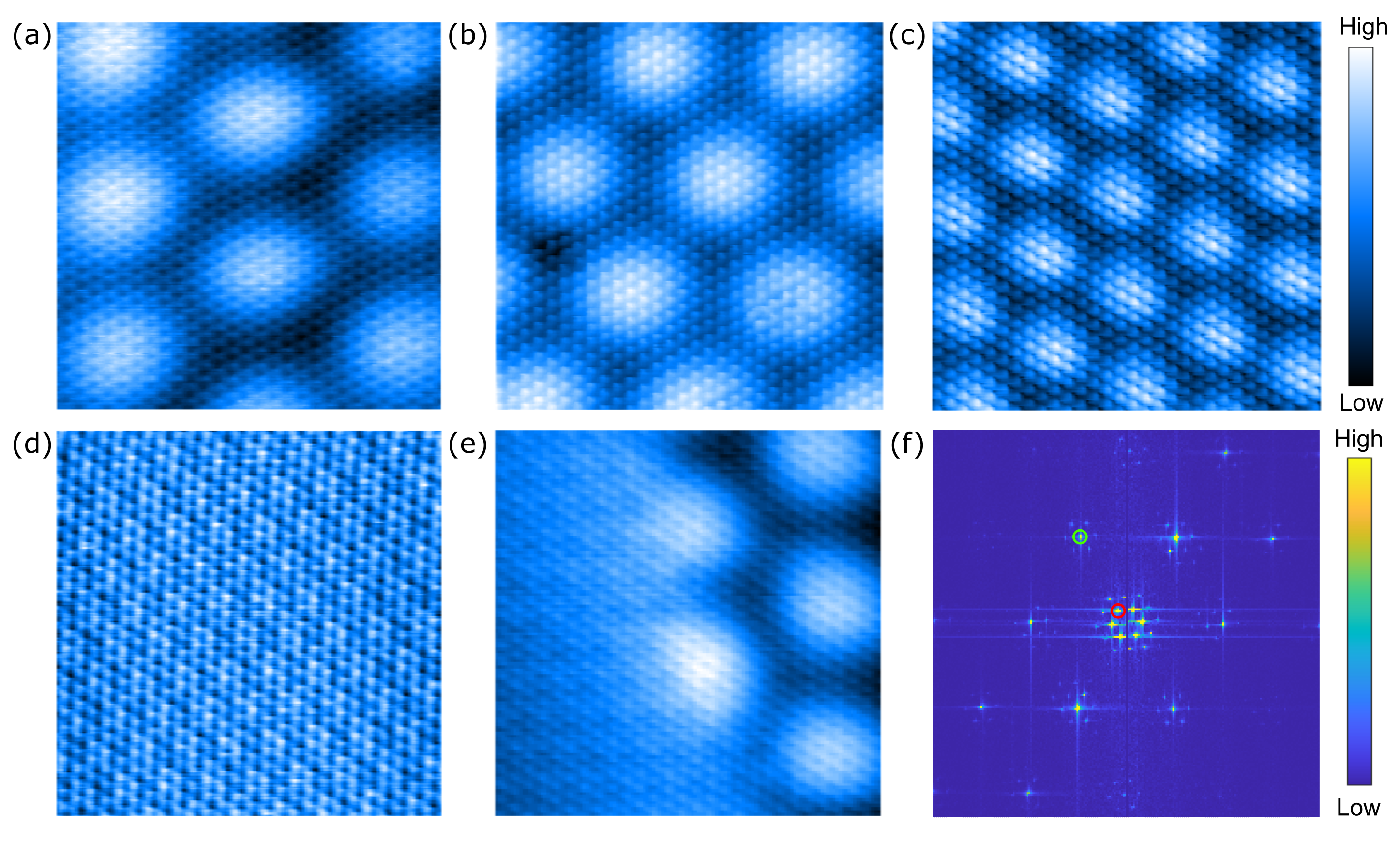}
    \caption{8$\times$8~nm$^{2}$ high resolution STM topography of MoS$_2$ on Au(111) with a twist angle of (a) 0.5\textdegree~(setpoint I$_t$= 200~pA; V$_b$= 500~mV), (b) 2.2\textdegree~(500~pA; 100~mV), (c) 7.7\textdegree~(200~pA; 300~mV) and (d) 21\textdegree~(100~pA; 100~mV). (e) Domain boundary between a 21\textdegree~and a 0.5\textdegree~twist angle region (100~pA; 1~V). (f) FT of (b) with the MoS$_2$ lattice and the moir\'e peaks identified with a green and a red circle, respectively.}
    \label{fig:figure1}
\end{figure*}

Exfoliating large area MoS$_2$ MLs onto poly-crystalline Au(111) naturally produces regions with different twist angles between the two lattices. This provides a unique platform to characterize the electronic structure for different twist angles in a single sample, thus excluding potential sample-dependent fluctuations. Figure~\ref{fig:figure1} shows typical raw data topographic STM images acquired in different regions of such a sample. They reveal the MoS$_2$ lattice and the moir\'e patterns specific to the local twist angle (Figure~\ref{fig:figure1}(a)-(d) and SI Section~\Romannum{2}). The corresponding Bragg and moir\'e peaks are highlighted in the Fourier transform (FT) of the 7.7\textdegree~heterostructure by a green and a red circle, respectively  (Figure~\ref{fig:figure1}(f)). The complete data set of Figure~\ref{fig:figure1} is consistent with a large single crystal MoS$_2$ ML, with a lattice oriented along a single direction, including across step edges and across gold grain boundaries (Figure~\ref{fig:figure1}(e) and SI Section \Romannum{3}).  

\begin{figure}[htp]

    \includegraphics[width=\columnwidth]{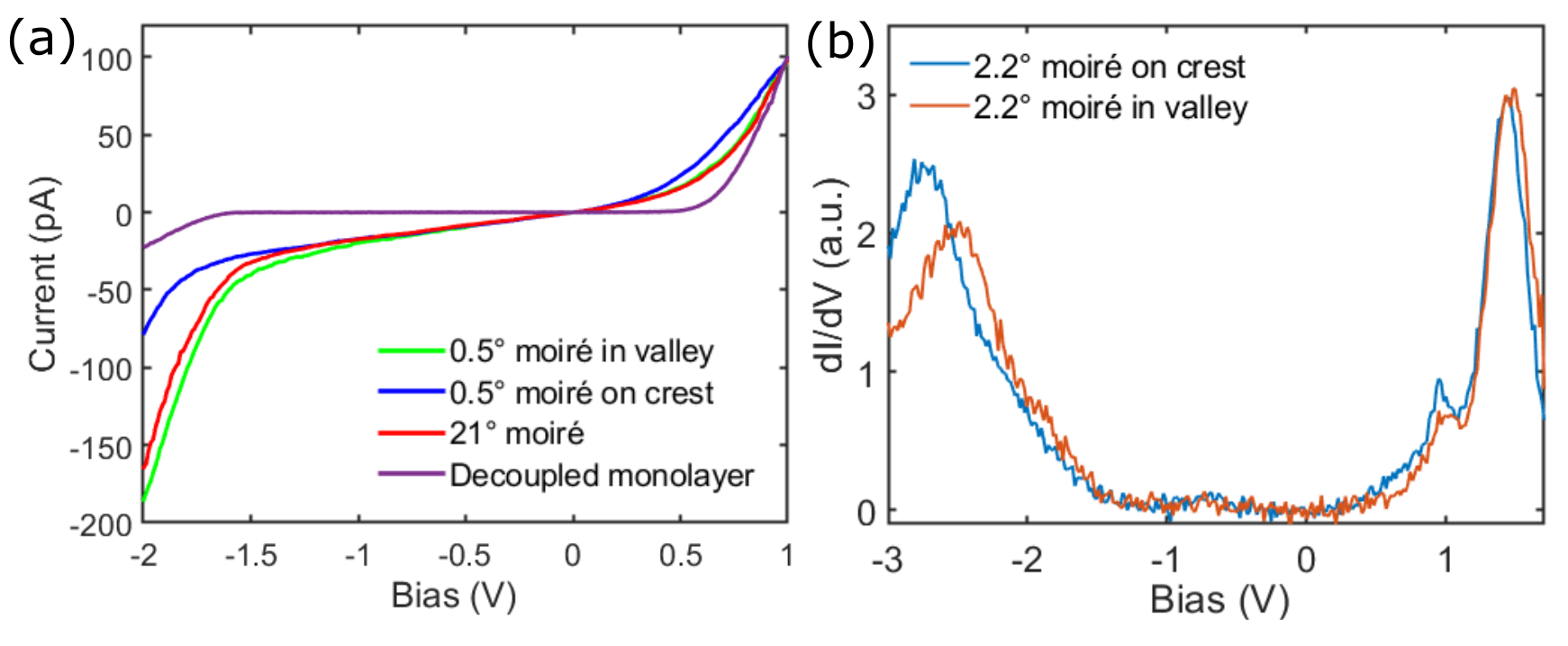}
    \caption{(a) $I(V)$ tunneling spectra measured at a valley (green) and at a crest (blue) location of a 0.5\textdegree~moiré pattern, on a 21\textdegree~moiré pattern (red), and in a decoupled region of the MoS$_2$ ML (purple). (b) $dI/dV(V)$ spectra measured at a crest (blue) and at a valley (orange) location of a 2.2\textdegree~twist angle moir\'e pattern.}
    \label{fig:figure2_new}
\end{figure}

We now turn to the spectroscopic characterization of MoS$_2$ on Au(111) as a function of twist angle. Typical $I(V)$ spectra measured on our devices are shown in Figure~\ref{fig:figure2_new}(a). They can be described as a modified semiconducting spectrum with a finite conductance in the gap region \cite{Blue2020, Kerelsky2017}. We do not observe the spread in tunneling characteristics reported in previous STM experiments \cite{Blue2020}, most likely due to a cleaner MoS$_2$/Au(111) interface exemplified by the perfect adherence of MoS$_2$ to the substrate across grain boundaries and step edges (SI Section \Romannum{3}). We do find occasional bubbles \cite{Wu2020, Peto2019}, where MoS$_2$ is locally decoupled from the substrate (SI Section \Romannum{4}). These regions are electronically different, with a fully gaped semiconducting $I(V)$ spectrum (Figure~\ref{fig:figure2_new}(a)) consistent with previous STM studies where MoS$_2$ was not in direct contact with a metal surface \cite{Peto2019, Zhou2016, Krane2016, Ponomarev2018}. 
 
\begin{figure*}[htp]

    \includegraphics{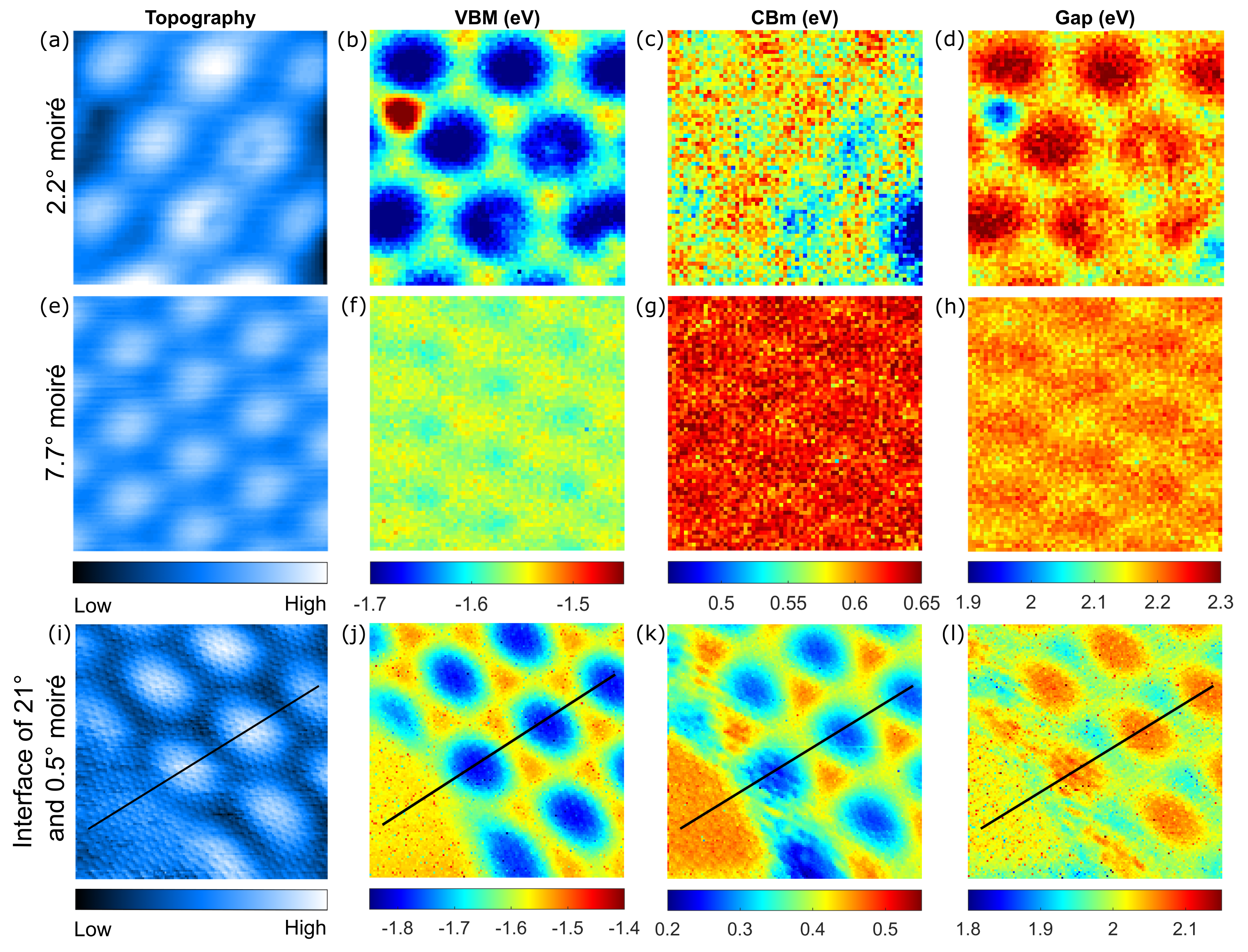}
    \caption{(a) 8$\times$8~nm$^2$ STM topography of a 2.2\textdegree~twist angle heterostructure (set point: I$_t$=100~pA, V$_b$=1.7~V) and corresponding (b) VBM, (c) CBm, and (d) gap map. (e) 6$\times$6~nm$^2$ STM topography of a 7.7\textdegree~twist angle heterostructure (100~pA, 1.0~V) and corresponding (f) VBM, (g) CBm, and (h) gap map -- the color scales for both twist angles are at the bottom of the second row; set-point for the $dI/dV(V,\Vec{r})$ maps was I$_t$=100 pA and V$_b$=1.7 V. (i) 9$\times$9~nm$^2$ STM topography of an interface between a 21\textdegree~(bottom left corner) and a 0.5\textdegree~twist angle heterostructure (100~pA, 1.0~V) and corresponding (j) VBM, (k) CBm, and (l) gap map (extracted from $I(V,\vec{r})$ spectra acquired with set point: 100 pA, 1.0~V).}
    \label{fig:figure2}
\end{figure*}
 
\begin{figure}[htp]

         \includegraphics[width=\columnwidth]{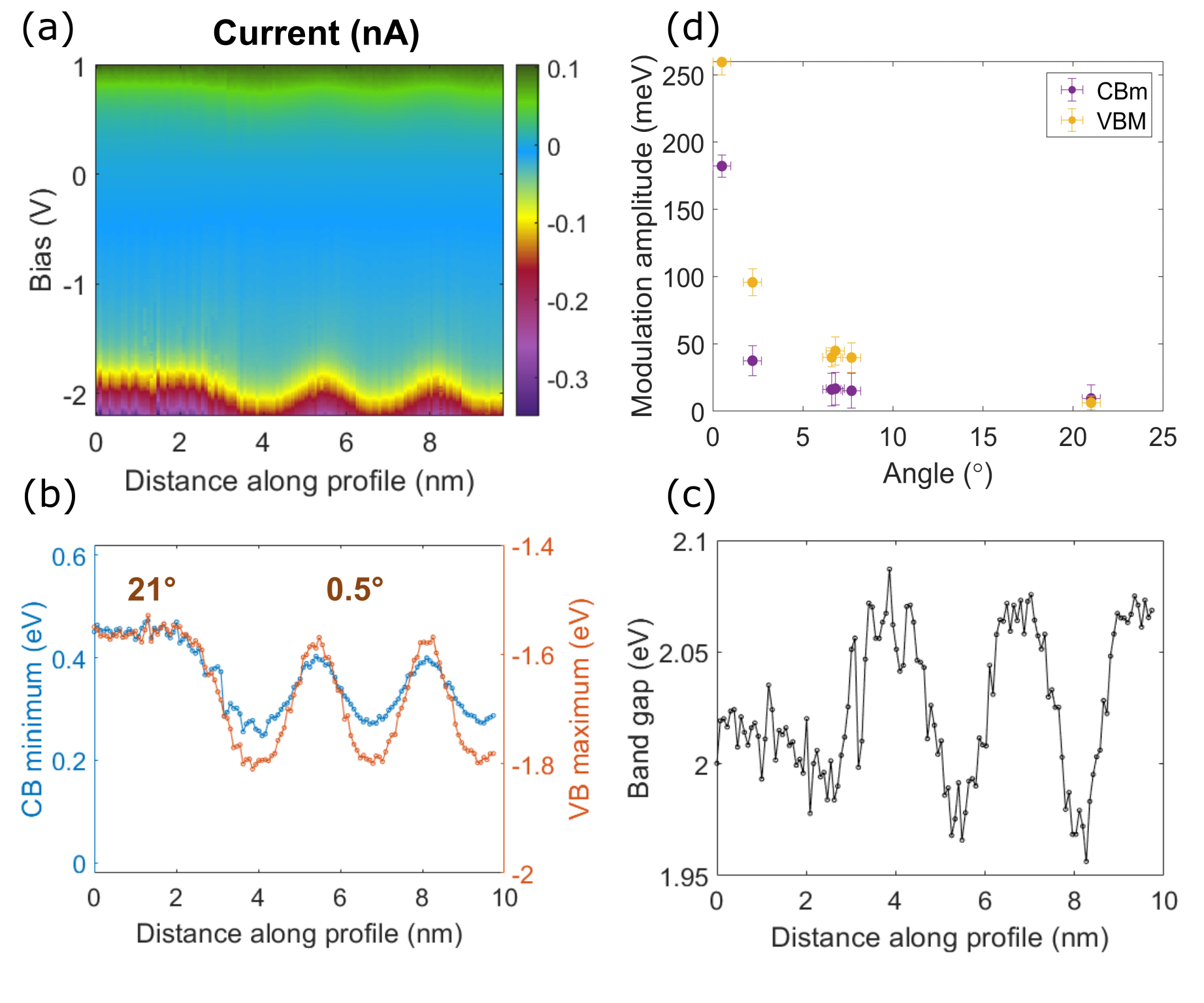}
         \caption{(a) $I(V,\vec{r})$ line-cut along the black lines in Figures~\ref{fig:figure2}(i)-(l), illustrating the spectroscopic evolution from a 21\textdegree~to a 0.5\textdegree~twist angle heterostructure. (b) Plots of the VBM, CBm, and (c) band gap as a function of position along the trace in (a). (d)  Modulation amplitude of the CBm and VBM as a function of twist angle.}
         \label{fig:figure5}
\end{figure}

While the generic line shape is the same for all $I(V)$ tunneling spectra measured on our MoS$_2$/Au(111) heterostructures, we find some variations, in particular as a function of twist angle and as a function of position in the moiré unit. They are most prominent at negative bias below the Fermi level ($E_F$), which corresponds to $V_\text{Bias}=0$~V in Figure~\ref{fig:figure2_new}. To characterize the twist angle dependence of the electronic properties of the MoS$_2$/Au(111) heterostructures, we acquire $I(V,\vec{r})$ and $dI/dV(V,\vec{r})$ maps. For every tunneling conductance spectrum --two typical examples are shown in Figure~\ref{fig:figure2_new}(b)--, we fit the main peaks below (above) $E_F$ to a Gaussian and define the valence band maximum $\text{VBM}(\vec{r})$ (conduction band minimum $\text{CBm}(\vec{r})$) as the peak position plus (minus) its corresponding 2.2$\sigma$. Since the conductivity measured at these energies is predominantly related to states derived from MoS$_2$ \cite{Bruix2016}, we define $\Delta (\vec{r})=\text{CBm}(\vec{r})-\text{VBM}(\vec{r})$ as the gap. We find that $\Delta$ amounts to about 2 eV in agreement with previous findings for MoS$_2$ MLs grown on Au(111) \cite{Krane2018, Silva2022}. These three quantities are plotted alongside the corresponding topography for two different twist angles in Figure~\ref{fig:figure2}(a)-(h). Note that the same information can be obtained from the $I(V,\Vec{r})$ curves using a different fitting procedure (SI section~\Romannum{5}) used in Figure~\ref{fig:figure2}(i)-(l) and in Figure~\ref{fig:figure5}(b)-(d). 

The three main results of our experiments can be graphically seen in Figure~\ref{fig:figure2}: i) the local density of states (DOS) is modulated at the moir\'e pattern wavelength; ii) the modulation amplitude is significantly larger for the valence band than for the conduction band, and so $\Delta(\vec{r})$ is also modulated at the moiré pattern wavelength; iii) the modulation amplitude of $\text{VBM}(\vec{r})$ and of $\text{CBm}(\vec{r})$ are decreasing with increasing twist angle. The vanishing spatial modulations of the band edges and of the gap with increasing twist angle are most strikingly seen in Figure~\ref{fig:figure2}(i)-(l) and in Figure~\ref{fig:figure5}. They show a domain wall between a 21\textdegree~and a 0.5\textdegree~twist angle region spanned by a single MoS$_2$ ML, which completely excludes any other origin than the twist angle for the observed differences. In Figure~\ref{fig:figure5}(b), we plot the modulation amplitudes of $\text{VBM}(\vec{r})$ and $\text{CBm}(\vec{r})$ --defined as the difference between maximum and minimum band edge energies for each twist angle-- as a function of twist angle. The plot clearly shows the monotonic reduction of the $\text{VBM}(\vec{r})$ and $\text{CBm}(\vec{r})$ modulation amplitudes with increasing twist angle.

\section{Discussion}

Density functional theory (DFT) calculations show that hybridization of MoS$_2$ with Au(111) happens primarily through sulfur $p$-orbitals and gold $d$-orbitals, and that it is strongest when the two atoms are positioned atop each other (on-top alignment) \cite{Reidy2021}. Calculations further show that the hybridization mainly involves out-of-plane orbitals --which define the valence band (VB)-- and less in-plane orbitals --which define the conduction band (CB)-- \cite{Bruix2016}. This is consistent with the greater modulation of the VBM compared to the CBm in Figure~\ref{fig:figure2} and in Figure~\ref{fig:figure5}(a)-(c). The modulation of the gap at the moir\'e periodicity in Figure~\ref{fig:figure2}(d),(h),(l) and in Figure~\ref{fig:figure5}(d) is a direct consequence of the different responses of the CB and of the VB to the hybridization. The effect is strongest in the middles of the bright moiré maxima seen in the topographic images in Figures~\ref{fig:figure1} and \ref{fig:figure2}. These moir\'e maxima must therefore correspond to the on-top alignment positions in the heterostructures, in agreement with previous assessments \cite{Krane2018, Silva2022}. 

Constant current STM images are a convolution of morphologic and electronic features \cite{chen}. Therefore, it is not a priori clear whether the moir\'e superstructure observed by STM is a structural or an electronic modulation. To address this question, we examined $dI/dV(V,\vec{r})$ conductance maps as a function of bias. The fact that the moir\'e contrast does fully invert for certain bias ranges (SI Section~\Romannum{6}) is not compatible with a structural modulation and provides strong evidence that the moir\'e pattern is of electronic origin in a flat MoS$_2$ ML, in agreement with STM and X-ray standing wave measurement by Silva \textit{et al.} \cite{Silva2022}.

\begin{figure}[htp]

          \includegraphics[width=\columnwidth]{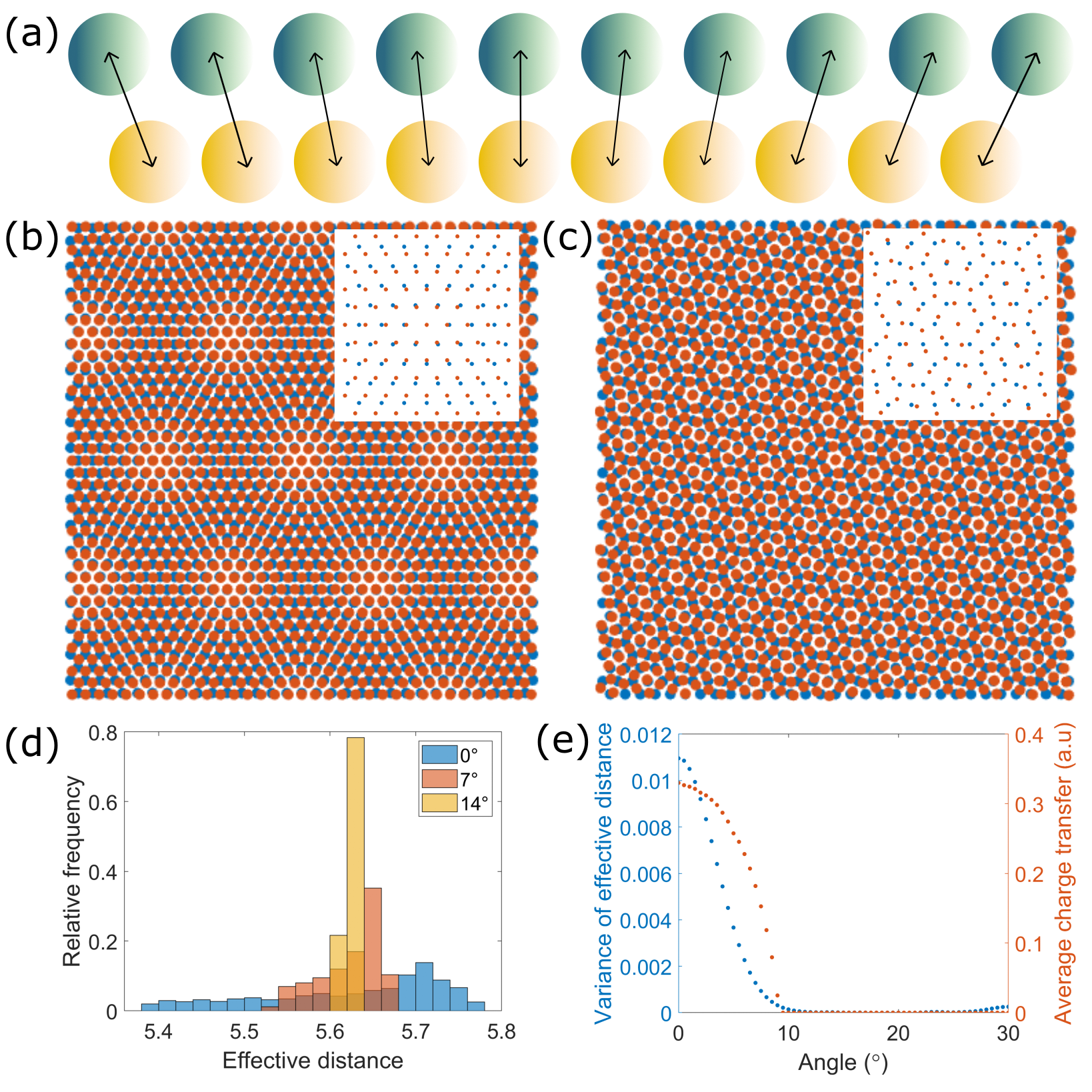}
          \caption{(a) Schematic side view of the bottom sulfur (green) on top Au (yellow) layers of the heterostructures. Schematic top view of a (b) 0\textdegree~and and a (c) 20\textdegree~twist angle moir\'e pattern with Au atoms in blue and S atoms in red. (d) Histogram of the effective distances $d_\text{eff}$ in a 0\textdegree, 7\textdegree~and 14\textdegree~twist angle heterostructure. (e) Twist angle dependence of the variance of the effective S-Au distance ($\sigma^2(d_\text{eff})$, left blue axis) and of the average charge transfer ($\overline{\Delta Q}$, right orange axis).}
          \label{fig:figure6}
      \end{figure}

Considering a flat MoS$_2$ ML on Au(111), we construct a simple model to understand the twist angle dependence of the electronic structure. We assume that the hybridization is primarily determined by the nearest neighbor distances (NND) between S and Au atoms shown as green and yellow spheres in Figure~\ref{fig:figure6}(a), respectively. The schematic top views in Figure~\ref{fig:figure6}(b) and (c) illustrate the changing registry of these two atomic layers as a function of twist angle and the resulting moir\'e patterns with spatially modulated NND. The distribution of the distances between the nearest S and Au atoms is independent of the twist angle (see inset in Figure~\ref{fig:figure6}(b) and (c)). In particular, the number of S atoms sitting on top of a Au atom per unit area is the same for all twist angles. Based on the registry between S and Au atoms alone, one would thus not expect any twist angle dependence of the amplitude of the modulated hybridization, in contradiction with the experimental observation. However, looking only at the NND to quantify the hybridization is not physically plausible: it neglects screening which prevents abrupt changes in the charge distribution over very short distances \cite{Zhang2018}. To take this into account we, introduce an effective distance $d_{\text{eff}}$ obtained by convoluting the spatial distribution of NNDs with a Gaussian. By construction, $d_{\text{eff}}$ is spatially modulated at the moir\'e period and provides a measure of the strength of the local hybridization: it is stronger where $d_{\text{eff}}$ is smaller. 

While the distribution of the distances between nearest S and Au atoms does not depend on twist angle, their spatial distribution does, with significantly larger site-to-site variations at larger twist angles as seen in Figure~\ref{fig:figure6}(b) and (c). These abrupt changes are attenuated by screening, leading to an increasingly narrow distribution of $d_{\text{eff}}$, as shown by the histograms in Figure~\ref{fig:figure6}(d) for 0\textdegree, 7\textdegree, and 14\textdegree~twist angles, which correspond to an increasingly homogeneous $d_{\text{eff}}$. To assess the amplitude of the modulated hybridization, we plot the variance of $d_{\text{eff}}$ (which reflects the width of its distribution) as a function of twist angle in Figure~\ref{fig:figure6}(e). This clearly shows that the spatial variation of the hybridization vanishes with increasing twist angle, which explains the correlation between large twist angles and more homogeneous electronic properties observed in Figures~\ref{fig:figure2} and \ref{fig:figure5}.

The average of the VBM and the average of the CBm both shift down in energy, with a larger shift for smaller twist angles (Figure~\ref{fig:figure5}(c)). This indicates a stronger electron doping with a larger overall charge transfer when the twist angle is small. The amount of charge transferred at a given site $i$ is a function of the local $d_{\text{eff}}(\vec{r}_i)$: $\Delta Q_i=f(d_{\text{eff}}(\vec{r}_i))$. The average charge transfer (i.e. charge transfer per Au-S pair for $N$ pairs) is
\begin{equation}
\overline{\Delta Q}=\frac{1}{N}\sum_{i=1}^{N}\Delta Q_i=\frac{1}{N}\sum_{i=1}^{N}f(d_{\text{eff}}(\vec{r}_i)).
\end{equation}

To gain insight into the twist angle dependence of $\overline{\Delta Q}$ within our simple model, we note that although the distribution of $d_{\text{eff}}$ depends on twist angle, its spatial average is constant (SI Section~\Romannum{7}). It means that $f(d_{\text{eff}}(\vec{r}_i))$ is not a linear function. 
We consider two simple non-linear $f(d_{\text{eff}}(\vec{r}_i))$ models which reproduce a stronger charge transfer for the Au-S pairs where $d_{\text{eff}}$ is shorter. Both models yield the same qualitative result. 

The first model is motivated by the finding of Silva et al. \cite{Silva2022} in MBE-grown films (0.5\textdegree~twist angle) that all the bottom S atoms of the MoS$_2$ ML fall into either of two categories: strongly bound or weakly bound to the underlying Au atom. Thus we consider two kinds of S atoms, one that contributes significantly (the strongly bound) and one that does not contribute to the charge transfer (the weakly bound). Assuming a given fraction (e.g. \(\frac{1}{3}\)) of significantly contributing S atoms in the 0\textdegree~twist angle heterostructure, we can estimate a general cut-off $d_{\text{eff,c}}$ below (above) which all S-Au pairs fall in the strongly (weakly) bound category with significant (negligible) charge transfer at a given twist angle. In this case 
\begin{equation}
f(d_\text{eff}(\vec{r}_i))=\begin{cases}
1 &\text{if $d_{\text{eff}}(\vec{r}_i)<d_\text{eff,c}$}\\
0 &\text{if $d_{\text{eff}}(\vec{r}_i)>d_\text{eff,c}$}
\end{cases}
\label{eq:model_1}
\end{equation}
In Figure~\ref{fig:figure6}(e) (right orange axis) we show $\overline{\Delta Q}$ as a function of twist angle calculated using Equation~\ref{eq:model_1}. We find that  $\overline{\Delta Q}$ monotonically decreases as a function of the twist angle, which reproduces the observed overall decreasing charge transfer with increasing twist angle.
We obtain the same result with another non-linear function $f(d_\text{eff}(\vec{r}_i))=1/d_{\text{eff}}$, as shown in SI Section~\Romannum{7}.

\section{Conclusions}
We present a systematic study of the twist-angle dependent electronic properties of exfoliated MoS$_2$ monolayers on Au(111) using high-resolution scanning tunneling microscopy and spectroscopy. We find that the conduction and valence bands are modulated at the moir\'e pattern period. The modulations are most prominent in the valence band and largest for the smaller twist angles. They vanish with increasing twist angles. We propose a simple model to understand this twist angle dependence based on a changing hybridization between S and Au orbitals, which does not only depend on the relative positions of the nearest S and Au atoms, but also on their neighboring configurations. These findings provide detailed insight into designing monolayer-on-metal heterostructures with variable electronic properties and doping by adjusting the twist angle. They also provide a platform to explore correlated and ordered electronic phases in combination with periodic charge transfer (or doping) patterns. 

\section{Methods}

We use gold-assisted exfoliation \cite{Magda2015, Huang2020, velicky2018} onto template-stripped gold substrates \cite{Hegner1993,Vogel2012} to mechanically isolate MoS$_2$ monolayers (MLs). The gold substrates are prepared in-house. First, we evaporate gold onto a clean ultra-flat silicon (Si) wafer. Second, we epoxy another flat Si piece onto the crystalline gold film. We then cleave this sandwich at the evaporated Au-Si interface to get an ultra-flat gold surface that reflects the flatness of the original Si substrate. Bulk 2H-MoS$_2$ single crystals were sourced from HQ graphene. These crystals are exfoliated to the monolayer limit onto the freshly exposed Au surface in a nitrogen-filled glove box, using scotch-tape exfoliation. The strong affinity between sulfur and the very clean and flat template-stripped Au substrate allows us to obtain millimeter-sized MoS$_2$ MLs. We identify ML MoS$_2$ flakes based on their optical contrast on gold and using Raman spectroscopy. A detailed characterization of the gold substrate and of the MoS$_2$-on-Au heterostructures is presented in the Supplementary Information Sections~\Romannum{1}~and~\Romannum{3}. Landing the STM tip on a desired region on the flake is performed using optical microscopy images. Throughout the process, we carefully protect the samples from exposure to the ambient atmosphere to avoid contamination and device degradation. For Raman and optical measurements, the samples were placed in a customized air-tight container with optical access that can be sealed in a glove box. For transferring the samples from the glove box to the STM, we used a home-built vacuum suitcase which can be directly attached to the load-lock of the ultra-high vacuum STM chamber. Prior to the STM measurements, the samples were annealed in situ at 150\textdegree~C for about 100 hours to obtain an optimal surface.

All the scanning tunneling microscopy and spectroscopy experiments were done using a Specs JT Tyto STM at 77.7 K or 5 K, at a base pressure better than 1$\times10^{-10}$ mBar. We used electrochemically etched W or Ir tips, all carefully conditioned and characterized in situ on Au(111) single crystal. STM topographic images were recorded in constant current mode. $dI/dV(V)$ conductance curves were acquired using a lock-in with a sample bias modulation amplitude of 15 mV at 527 Hz. 

For the qualitative modeling of the twist angle-dependent hybridization, we considered a large number of atoms (of the order of $10^6$) in each layer to minimize finite size effects.

\section*{ACKNOWLEDGEMENTS}
We thank A.F. Morpurgo for providing access to glove box, evaporator, and AFM facilities. We thank A. Scarfato and I. Maggio-Aprile for stimulating scientific discussions, and A. Guipet and G. Manfrini for technical support in the STM laboratory. We also thank A. \O rsted and J.C. Crost for helping with the first attempts to prepare the Au substrates and J. Teyssier for his assistance with the Raman measurements. This work was supported by the Swiss National Science Foundation (Division II Grant No. 182652). 


 \bibliographystyle{ieeetr}
 \bibliography{biblio.bib}

\end{document}


\title{Supporting Information for 
Twist angle dependent electronic properties of exfoliated single layer MoS$_2$ on Au(111)}

\author{Ishita Pushkarna}
\affiliation{DQMP, Université de Genève, 24 Quai Ernest Ansermet, CH-1211 Geneva, Switzerland}
\affiliation{These authors contributed equally to this work.}
\author{\'{A}rp\'{a}d P\'{a}sztor}
\email{arpad.pasztor@unige.ch}
\affiliation{DQMP, Université de Genève, 24 Quai Ernest Ansermet, CH-1211 Geneva, Switzerland}
\affiliation{These authors contributed equally to this work.}
\author{Christoph Renner}
\email{christoph.renner@unige.ch}
\affiliation{DQMP, Université de Genève, 24 Quai Ernest Ansermet, CH-1211 Geneva, Switzerland}

\maketitle

\tableofcontents

\newpage
  
\section{Substrate and sample characterization}
\label{sup:sec_substrateandsample}
\begin{figure}[htp]
    \centering
    \includegraphics[width=\columnwidth]{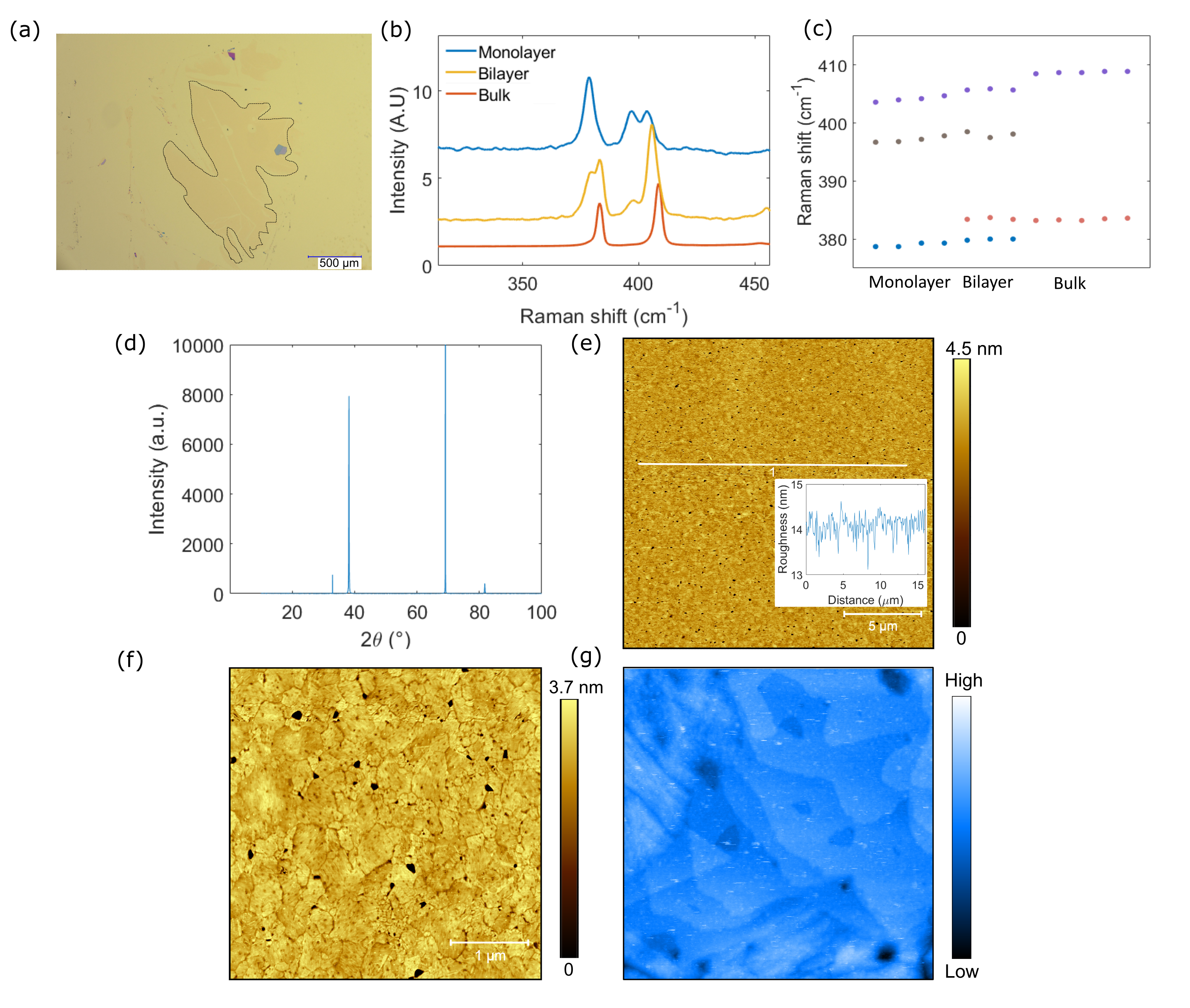}
    \caption{(a) Optical microscope image showing a millimeter-sized monolayer MoS$_2$ flake (black outline) on the gold substrate. (b) Raman spectra of monolayer (blue), bilayer (yellow), and bulk (orange) MoS$_2$ on Au. (c) Raman peak positions for three different MoS$_2$ flake thicknesses, with data taken at different positions on each flake. (d) XRD spectrum of the bare Au substrate, showing its (111) orientation (38.1\textdegree~and 81.81\textdegree~peaks). (e) 20$\times$20 $\mu$m$^2$ AFM image of the gold substrate, illustrating the ultra-flat surface over tens of microns. (f) 4$\times$4~$\mu$m$^2$ AFM image showing the polycrystalline nature of the substrate. (g) 0.2$\times$0.2 $\mu$m$^2$ STM image of the same substrate revealing gold terraces.}
    \label{fig:suppl_fig1}
\end{figure}

We obtained millimeter-sized monolayer (ML) MoS$_2$ by exfoliating 2H-MoS$_2$ single crystals onto template-stripped gold substrates. Exfoliation onto these substrates primarily resulted in large MLs, with only a few tiny thicker flakes. A typical ML is outlined in black in the optical microscope image in Figure~\ref{fig:suppl_fig1}(a). 

Exfoliated MLs were identified through the characteristic Raman spectra of MoS$_2$ on Au \cite{Velicky2020, Pollmann2021}. Due to the strong interaction with the substrate, ML MoS$_2$ has a markedly different Raman spectrum on gold (Figure~\ref{fig:suppl_fig1}(b), blue spectrum) than on SiO$_2$/Si \cite{Li2012}. The A$_{1g}$ and E$_{2g}$ modes in the ML are shifted with respect to their bulk positions on both substrates. However, the shift is different on Au, with an additional splitting of the A$_{1g}$ mode appearing around 397 cm$^{-1}$ (Figure~\ref{fig:suppl_fig1}(b), blue spectrum).
The bilayer spectrum features four distinguishable peaks (Figure~\ref{fig:suppl_fig1}(b), orange spectrum), which correspond to the combined peaks observed in bulk and ML specimen. This spectrum can be understood as a combined contribution from a bulk spectrum and from the first ML, which is affected by the substrate. The evolution of Raman peaks as a function of MoS$_2$ flake thickness is summarized in Figure~\ref{fig:suppl_fig1}(c) where we show the position of the peaks at several different locations on each flake. 

The gold substrates were characterized using X-ray diffraction (XRD) and atomic force microscopy (AFM). The XRD spectrum in Figure~\ref{fig:suppl_fig1}(d) shows a peak around 38.1\textdegree, indicating (111) orientation, and a peak near 81.81\textdegree~corresponding to the (222) reflection of the gold surface. The peaks at 34\textdegree~and 69.25\textdegree~correspond to the Si(100) substrate \cite{Krishnamurthy2014,Zhao2004}. 
The large-scale AFM image in Figure~\ref{fig:suppl_fig1}(e) reveals an ultra-flat gold surface, with a typical roughness of less than a nanometer. Disposing of such flat and freshly exposed gold surfaces is essential to exfoliate large-area MoS$_2$ MLs. The polycrystalline structure of the gold substrate is nicely resolved in smaller range AFM (Figure~\ref{fig:suppl_fig1}(f)) and STM images (Figure~\ref{fig:suppl_fig1}(g)).

\newpage
\clearpage

\section{Determination of the twist angle}
\label{sup:sec_twistangle}

\begin{figure}[ht]
    \includegraphics{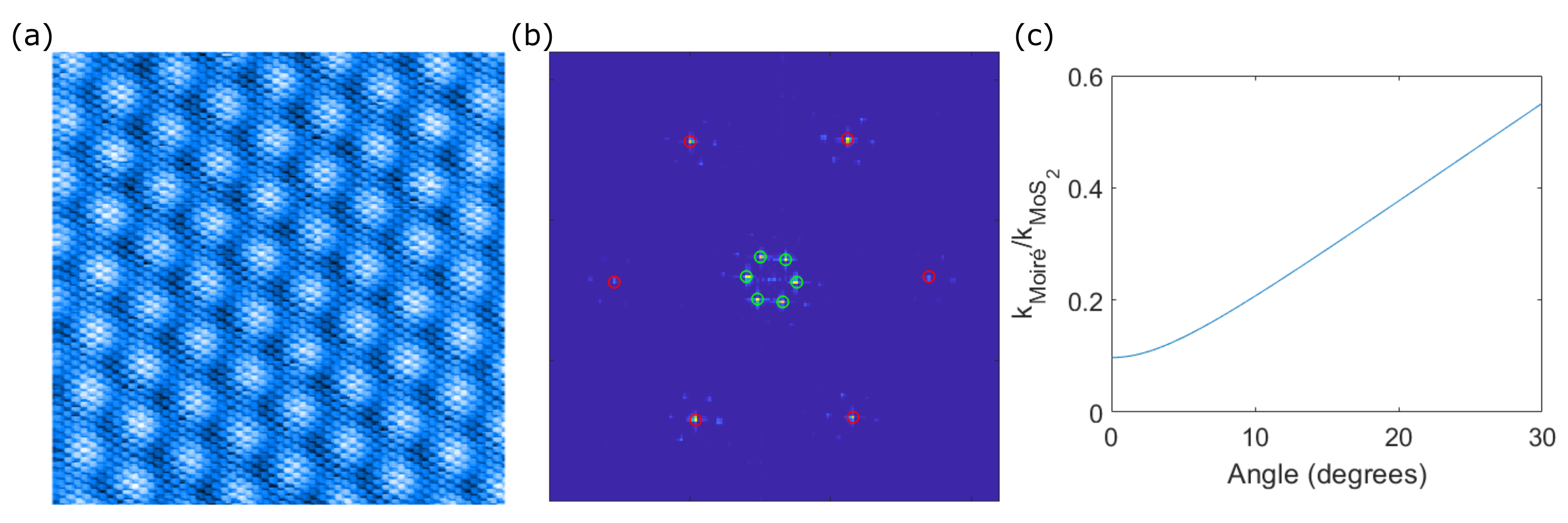}
    \caption{(a) 15$\times$15~nm$^2$ STM topography of a MoS$_2$ ML on Au(111) aquired at 300~mV and 100~pA set point. (b) Corresponding FT with the MoS$_2$ lattice (red) and moir\'e (green) peaks selected to determine the twist angle (7.7\textdegree~in this case). (c) Plot of $|\vec{k}_\text{Moiré}|/|\vec{k}_{\text{MoS}_2}|$ as a function of twist angle.}
    \label{fig:suppl_fig2}
\end{figure}

We determine the twist angle between the MoS$_2$ lattice and the Au(111) surface based on the Fourier-transform (FT) of atomically resolved STM topographies exemplified in Figure~\ref{fig:suppl_fig2}(a). We identify and select the peaks corresponding to the moiré and to the MoS$_2$ wave vectors marked by green and red circles in Figure~\ref{fig:suppl_fig2}(b), respectively, and calculate the length of each wave vector. The length of the gold lattice wave vector is given by the sum of the wave vectors of MoS$_2$ and moiré lattices. We choose the pairs of moiré and MoS$_2$ wave vectors which give the value closest to the lattice constant of gold to calculate the corresponding twist angle using supplementary equation~(1). In the example of Figure~\ref{fig:suppl_fig2}, we find a twist angle of 7.7\textdegree. The uncertainty on the twist angles determined in this way is $\pm 0.5$\textdegree, primarily limited by the precision of measuring the k-vectors in the FTs. 
\begin{equation}
    \begin{split}
    cos({\varphi})&=\frac{\vec{k}_{\text{Au}}\cdot\vec{k}_{\text{MoS}_2}}{|\vec{k}_{\text{Au}}||\vec{k}_{\text{MoS}_2}|}
    \end{split}
\end{equation}

An alternative approach to determine the twist angle is to calculate the moiré wavelength as a function of the twist angle. This allows to extract the twist angle from the plot of $|\vec{k}_\text{Moiré}|/|\vec{k}_{\text{MoS}_2}|$ as a function of twist angle (Figure~\ref{fig:suppl_fig2}(c)). 

\newpage
\clearpage

\section{Characterization of gold grains and their interfaces}
\label{sup:sec_goldstep}
\begin{figure}[htp]
    \centering
    \includegraphics[scale=0.95]{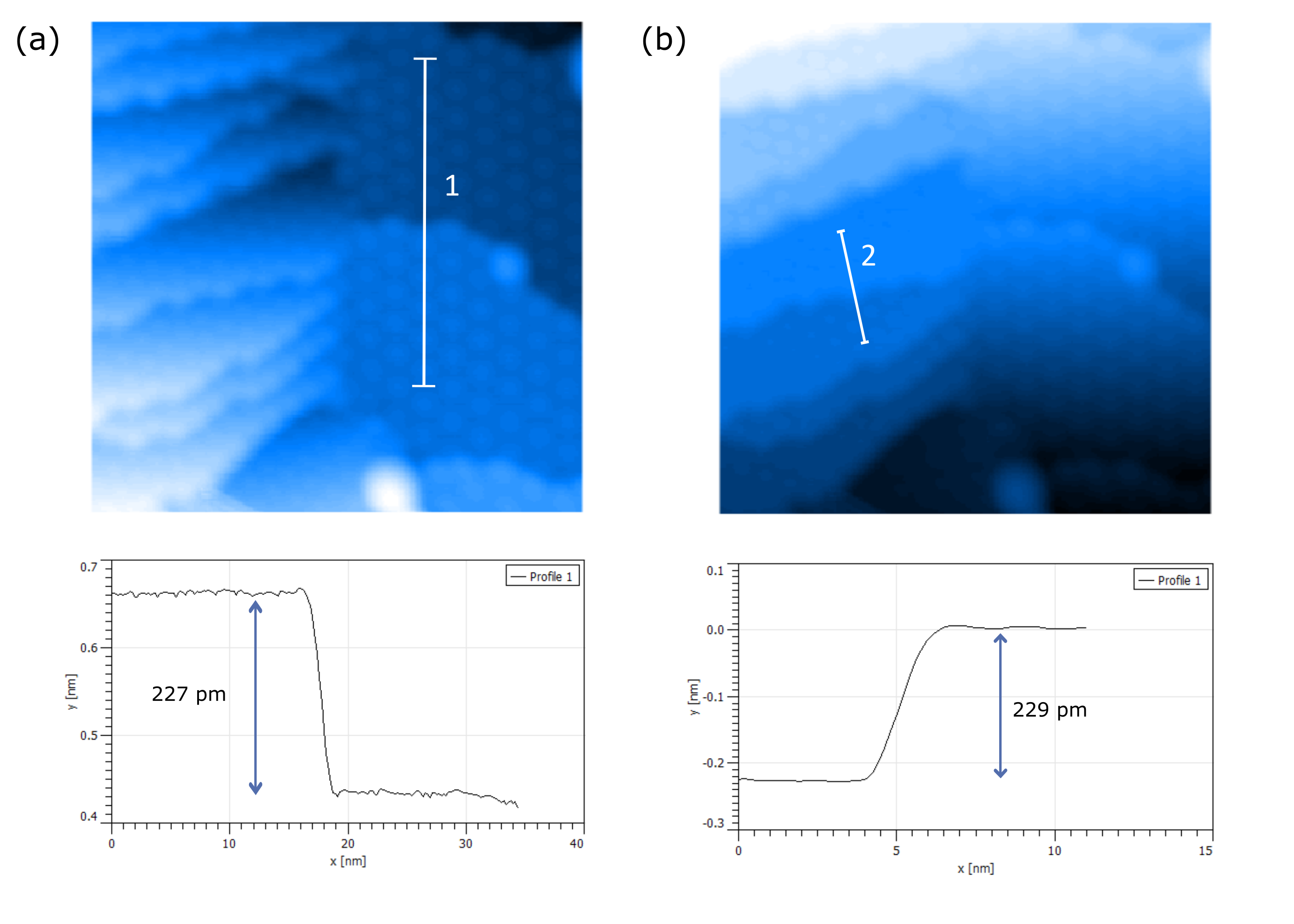}
    \caption{50$\times$50~nm$^2$ topographic STM image of two adjacent Au(111) grains leveled to show the terraces in the right-hand grain in (a) and the left-hand grain in (b), with corresponding height profile along the white line in each background corrected image.}
    \label{fig:suppl_fig8}
\end{figure}

XRD shows that the gold film is composed of [111]-oriented grains. High-resolution STM topography is consistent with XRD, and provides atomic-scale insight into the tilting and rotation of the grains about their [111]-axis. In Figure~\ref{fig:suppl_fig8}, we analyze a high-resolution topographic image of a continuous MoS$_2$ ML spanning two adjacent gold grains. We can determine the step heights in each grain by successively flattening the STM image with respect to a terrace on the left-hand side grain and then with respect to a terrace on the right-hand side grain. In both cases, we extract a height difference between the terraces of about 228~pm, consistent with the step height expected for Au(111) \cite{Barth1990, Sun2008}, as shown in the lower panels of Figure~\ref{fig:suppl_fig8}(a),(b). 

\begin{figure}[ht]
    \centering
    \includegraphics{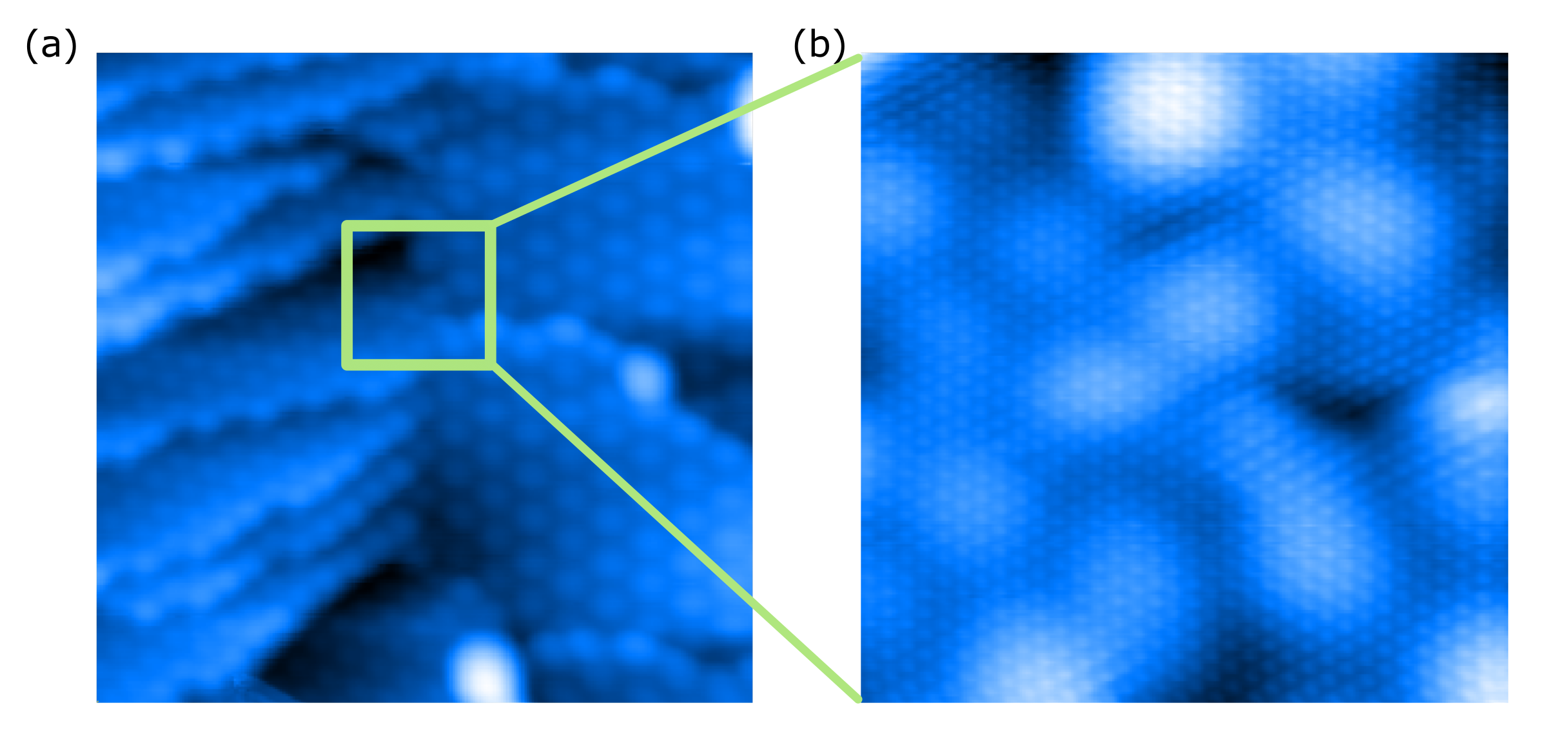}
    \caption{(a) 50$\times$50~nm$^2$ high resolution STM image (same as in Figure~\ref{fig:suppl_fig8}) levelled to highlight the moir\'e patterns. (b) Magnified 10$\times$10~nm$^2$ grain boundary region to emphasize the continuous MoS$_2$ lattice. The large balls correspond to the moiré pattern and the small ones to the MoS$_2$ atomic lattice.}
    \label{fig:suppl_fig4}
\end{figure}

The same topographic STM image as in Figure~\ref{fig:suppl_fig8} is shown in Figure~\ref{fig:suppl_fig4}(a), but leveled to highlight the moiré patterns in the two adjacent grains. A closer look at the grain boundary region in Figure~\ref{fig:suppl_fig4}(b) clearly shows a continuous MoS$_2$ flake extending over the entire field of view. The changing moir\'e pattern is thus a direct consequence of the different orientations of the two Au(111) grains. Continuous MoS$_2$ flakes spanning Au(111) grain boundaries provide a unique platform to study the electronic properties as a function of twist angle in a single device with the same tip, excluding any spurious experimental effects that might affect the data.

\newpage
\clearpage

\section{Nanobubbles on the surface}
\label{sup:sec_nanobubbles}
\begin{figure}[ht]
    \centering
    \includegraphics{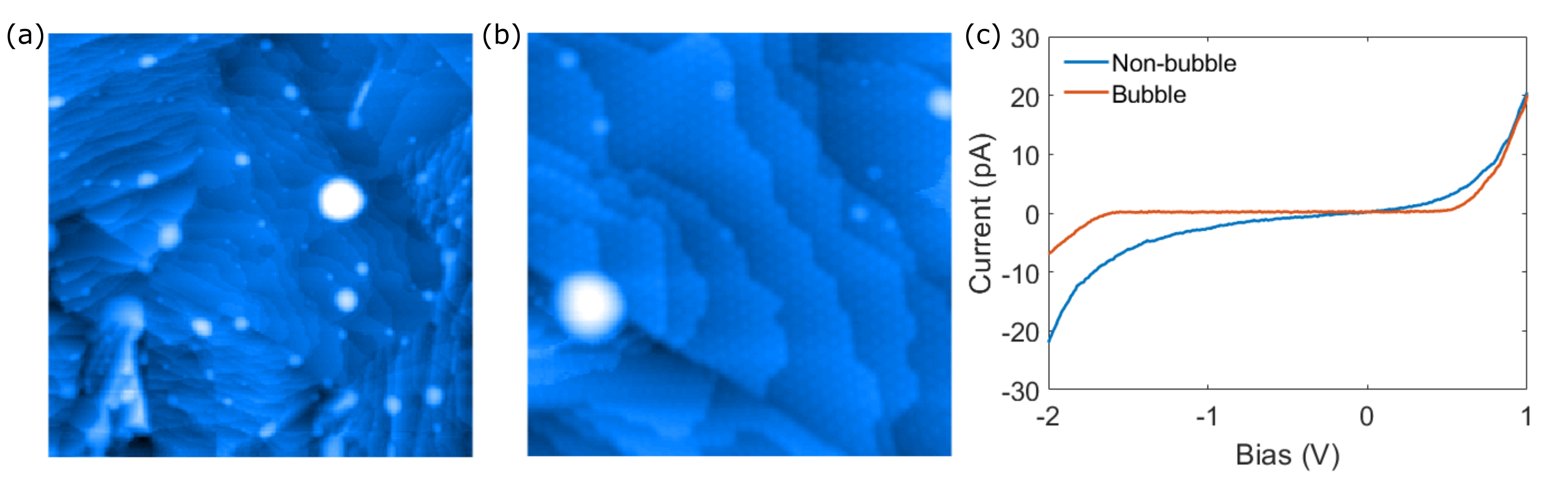}
    \caption{(a) 400$\times$400~nm$^2$ STM topography of a single continuous MoS$_2$ ML spanning different Au(111) grains and terraces. (b) Magnified 100$\times$100~nm$^2$ STM topography revealing the MoS$_2$ lattice and moir\'e patterns on every gold terrace where MoS$_2$ is in close contact with the Au(111) surface, and nanobubbles where MoS$_2$ is detached from the gold substrate. (c) Semi-conducting $I(V)$ spectrum measured on a bubble (orange), and a hybridized $I(V)$ spectrum measured off the bubble (blue).}
    \label{fig:suppl_fig3}
\end{figure}

The large area topographic STM image in Figure~\ref{fig:suppl_fig3}(a) shows a single continuous MoS$_2$ ML extending over the entire image and spanning different Au(111) grains and terraces. We observe occasional bubbles, usually located around step edges (Figure~\ref{fig:suppl_fig3}(b)). They correspond to regions where the MoS$_2$ ML is decoupled from the substrate. Similar features have been observed in other studies of exfoliated TMDs \cite{Peto2019} and graphene \cite{Khestanov2016} on different substrates. These decoupled regions show purely semi-conducting $I(V)$ spectra (Figure~\ref{fig:suppl_fig3}(c), orange spectrum), whereas the surrounding regions show hybridized spectra, with a reminiscence of the semi-conducting nature of MoS$_2$ (Figure~\ref{fig:suppl_fig3}(c), blue spectrum). 

\newpage
\clearpage

\section{Spatial mapping of VBM, CBm, and gap using $I(V, \vec{r})$ curves}
\label{sup:sec_IVfitting}
\begin{figure}[htp]
    \centering
    \includegraphics{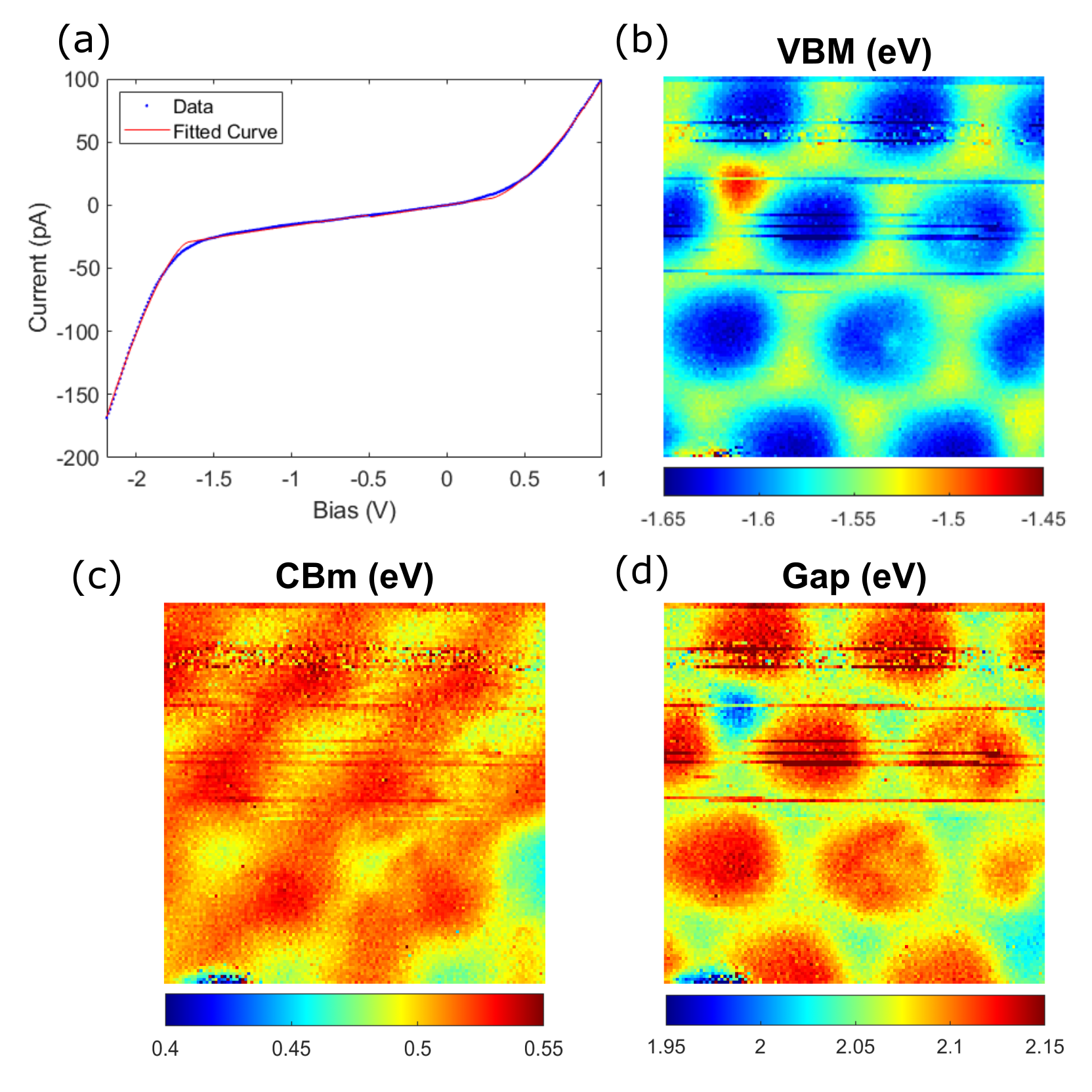}
    \caption{(a) $I(V)$ spectrum and corresponding fit. (b) VBM, (c) CBm, and (d) gap modulation of the 2.2\textdegree~moiré pattern extracted from a 8$\times$8~nm$^2$ I(V,$ \Vec{r}$) map acquired in the same area as Figure~3.}
    \label{fig:suppl_fig5}
\end{figure}

Here we show how we extract the CB and VB edges from $I(V)$ spectra by fitting the $I(V)$ curves following Zhuou et al. \cite{Zhou2016} with a slightly different model DOS. We considered a constant DOS for the purely semi-conducting decoupled MoS$_2$ areas. However, for the hybridized films, a better fit is obtained using a square-root energy-dependent semiconducting DOS (3D) in each band. To take into account the metallic background due to the hybridization, we added a non-zero constant to the DOS over the entire energy range. The VB and CB regions were then fitted separately, using independent constants for each half of the spectrum (each spectrum is divided into two parts at the approximate center of the gap near -0.5~V). This ensures a roughly equal number of data points to fit both sides. In Figure~\ref{fig:suppl_fig5}(a), we show an example of a fitted $I(V)$ spectrum.

Figure~\ref{fig:suppl_fig5}(b), (c), and (d) show the spatial mapping of VBM, CBm, and the band gap, obtained by fitting $I(V)$ curves. Both methods (fitting of $dI/dV(V)$ or $I(V)$ spectra) give the same information on band edge modulation and attest the suitability of any of the two methods.
\newpage
\clearpage

\section{Conductance maps ($dI/dV(V,\Vec{r})$)  as a function of bias }
\label{sup:sec_bias_dep_LDOS}

\begin{figure}[htp]
    \centering
    \includegraphics[scale=0.9]{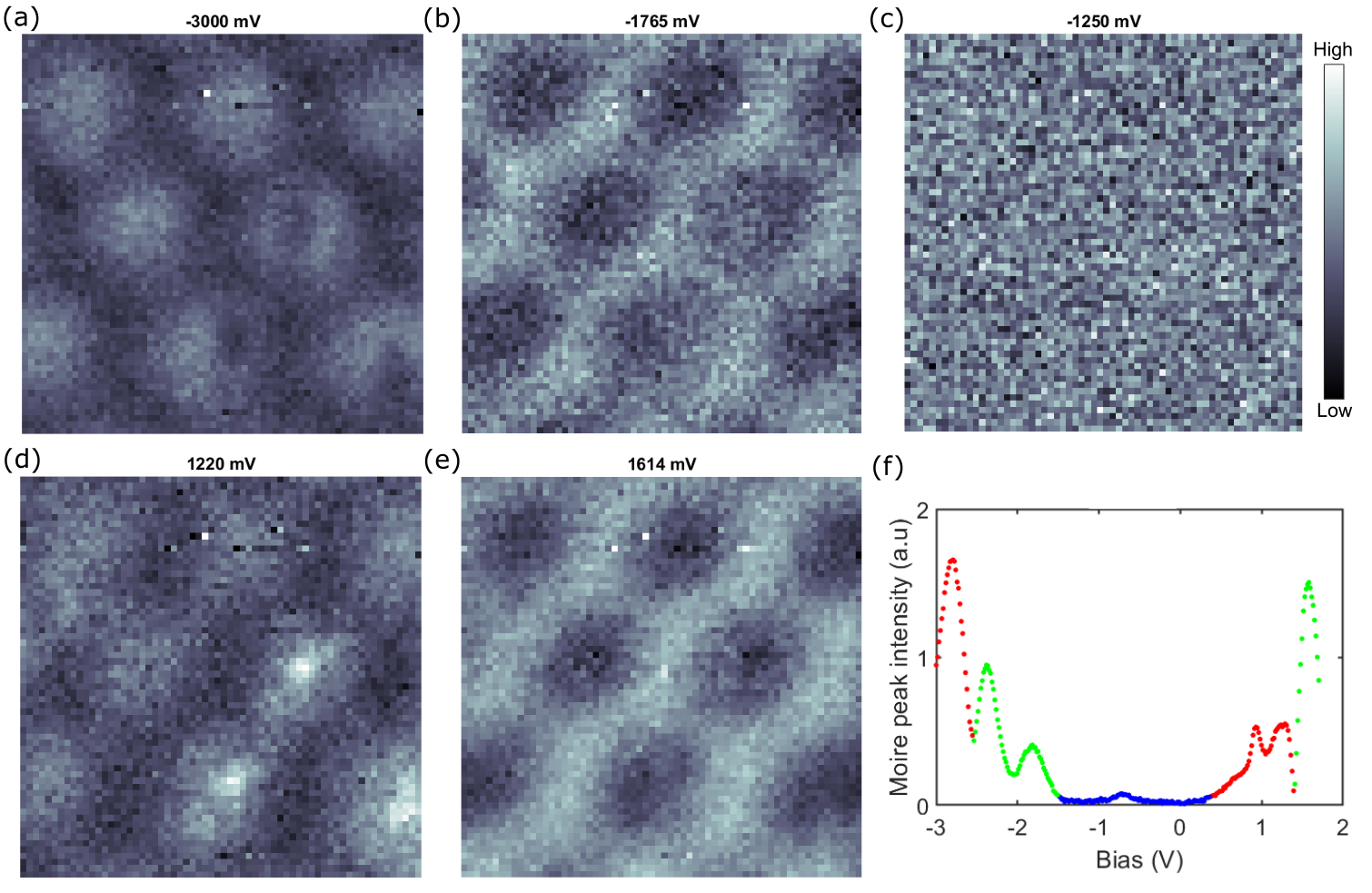}
    \caption{8$\times$8~nm$^2$ conductance maps extracted from a $dI/dV(V,\Vec{r})$ map of a 2.2\textdegree~twist angle heterostructure at (a) -3000~mV, (b) -1765~mV, (c) -1250~mV, (d) 1220~mV, and (e) 1614~mV. Set point during the conductance map: I$_t$= 100~pA; V$_b$= 1.7~V. (f) Plot of the total intensity of the six moiré peaks in the FT of all conductance maps available between -3~V and 1.7~V. Red corresponds to the contrast in panel (a), green to the contrast in panel (b), and blue to energies where the moiré peaks are not resolved in the FT.}
    \label{fig:figure4}
\end{figure}

To address the electronic or structural origin of the moiré pattern observed in STM images, we examined the $dI/dV(V,\Vec{r})$ conductance maps of a 2.2\textdegree~twist angle heterostructure as a function of bias in  Figure~\ref{fig:figure4}.  The main observations are the following: i) we do not see any significant modulation of the local density of states (LDOS) at bias voltages inside the MoS$_2$ gap (Figure~\ref{fig:figure4}(c)), probably because the tip is stabilized far outside the gap and the signal from these low energy states is too weak; ii) outside of the gap, the LDOS is modulated at the  moiré period (Figure~\ref{fig:figure4}(a),(b),(d),(e)); iii) the moiré contrast inverts at least three times in the examined energy range, once at negative bias (Figure~\ref{fig:figure4}(a)-(b)), once across the gap (Figure~\ref{fig:figure4}(b)-(d)), and once at positive bias (Figure~\ref{fig:figure4}(d)-(e)). Such contrast inversions in the conductance maps as a function of energy are not consistent with morphologic features: hills cannot swap their positions with trenches three times as a function of imaging bias. 

\begin{figure}[htp]
    \centering
    \includegraphics[scale=0.9]{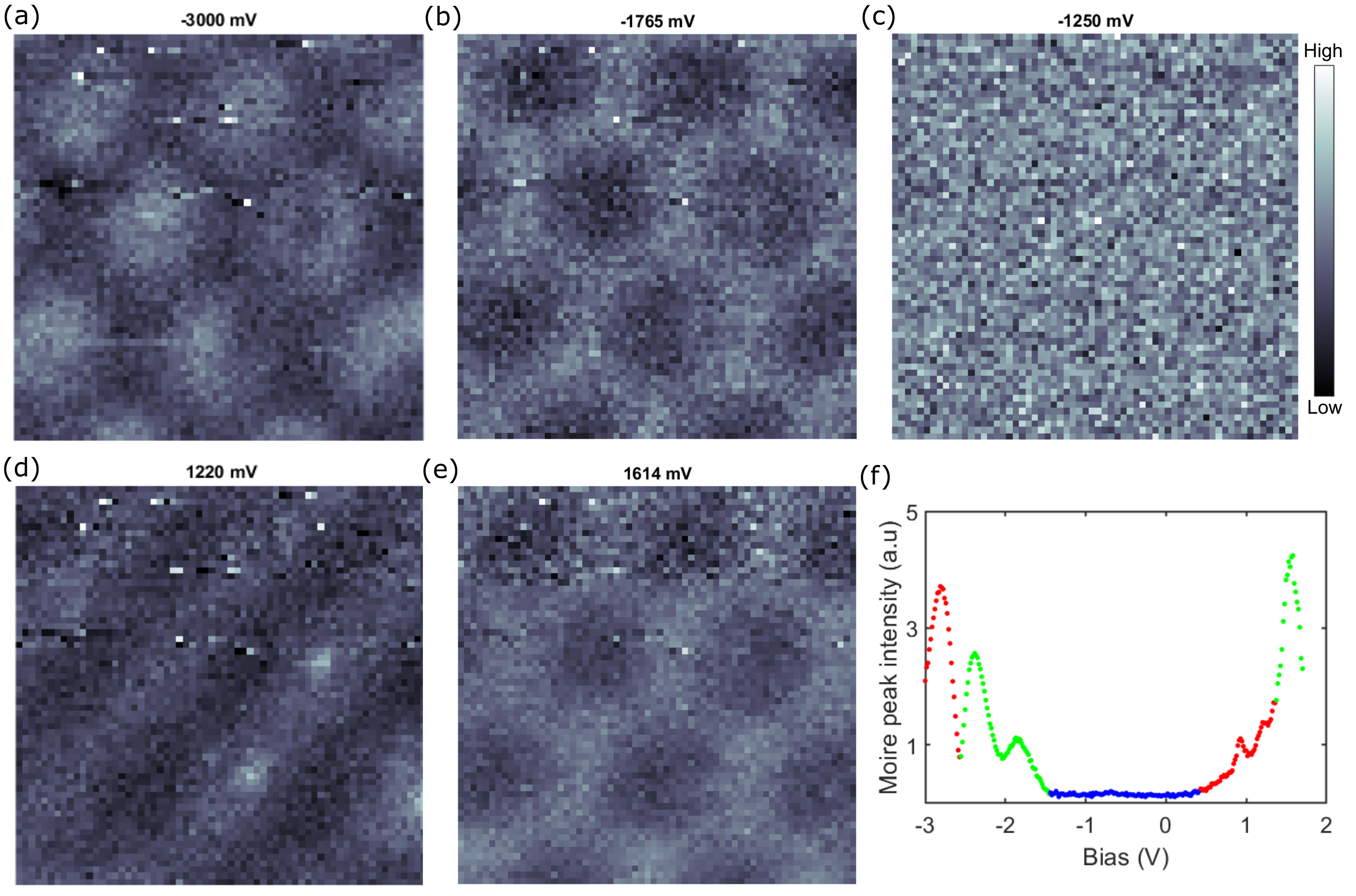}
    \caption{8$\times$8~nm$^2$ conductance maps extracted from a $dI/dV(V,\Vec{r})$ map of a 2.2\textdegree~twist angle heterostructure at (a) -3000~mV, (b) -1765~mV, (c) -1250~mV, (d) 1220~mV, and (e) 1614~mV. Set point during the conductance map: I$_t$= 100~pA; V$_b$= -3~V. (f) Plot of the total intensity of the six moiré peaks in the FT of all conductance maps available between -3~V and 1.7~V. Red corresponds to the contrast in panel (a), green to the contrast in panel (b), and blue to energies where the moiré peaks are not resolved in the FT.}
    \label{fig:suppl_fig6}
\end{figure}

\begin{figure}[htp]
    \centering
    \includegraphics{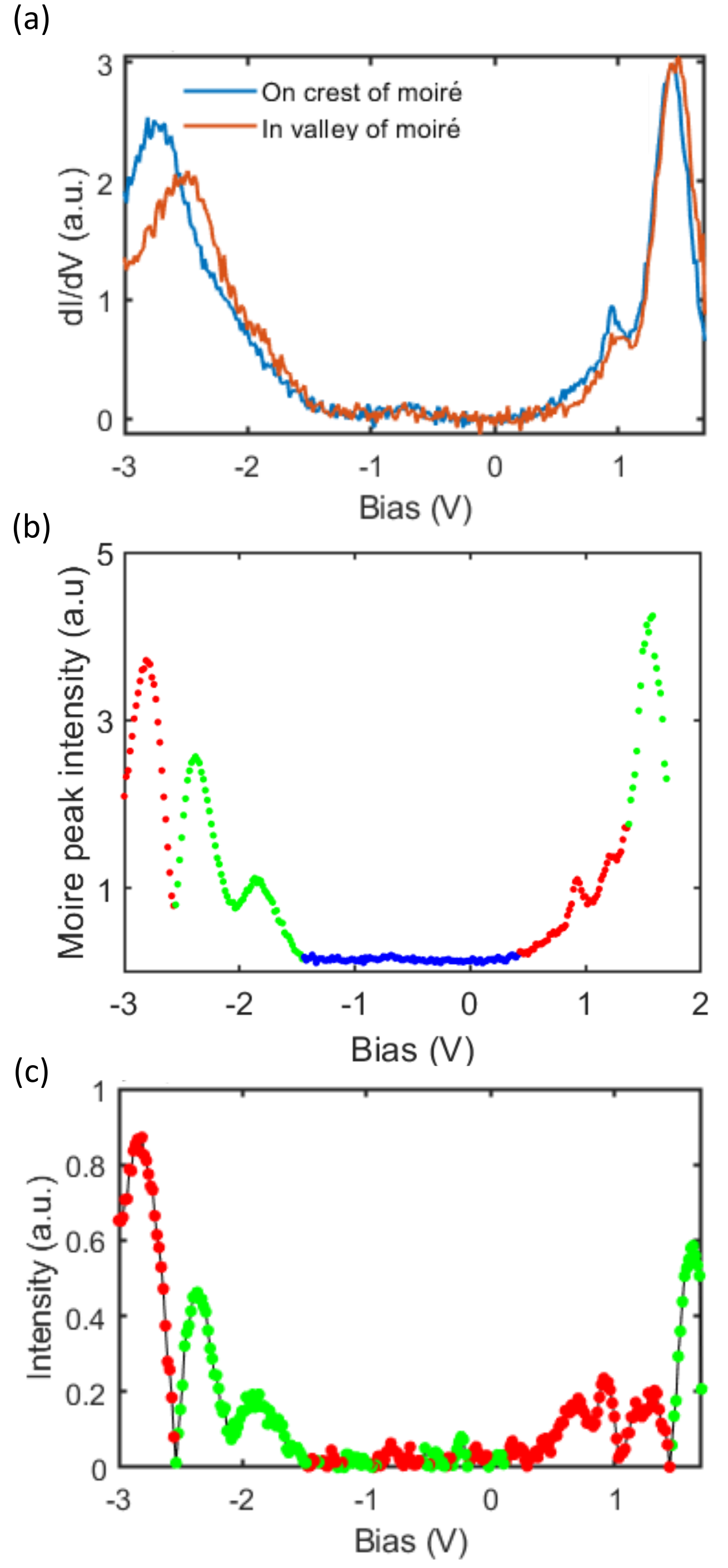}
    \caption{(a) $dI/dV(V)$ spectra measured at the crest (blue) and valley (orange) positions of the moiré pattern. (b) Moiré modulation amplitude extracted from the total intensity of the corresponding six peaks in the FT of the energy-dependent conductance maps measured on a 2.2\textdegree~twist angle heterostructure (same data as Figure~\ref{fig:suppl_fig6}(f)). (c) The absolute value of the \textit{on-crest} minus \textit{in-valley} $dI/dV(V)$ spectra in panel (a), where red (green) corresponds to a positive (negative) difference.}
    \label{fig:suppl_fig10}
\end{figure}

The electronic origin of the observed moiré pattern can be entirely understood by comparing the $dI/dV(V)$ curves measured at a crest and a valley position: they cross each other at several biases and are nearly equal in the gap region (see Figure~\ref{fig:suppl_fig10}(a) and Figure~2(b)). Therefore, depending on the bias chosen to extract the conductance maps, the LDOS will be largest when the tip is over a crest or when it is over a valley of the moiré, leading to changing and inverted contrasts. To make sure the bias set-point does not affect these observations, we did the same analysis for a $dI/dV(V,\Vec{r})$  map acquired at a negative bias set point of -3~V. The result is exactly the same as seen in Figure~\ref{fig:suppl_fig6}. 

The changing contrast of the moiré pattern in Figures~\ref{fig:figure4} and ~\ref{fig:suppl_fig6} is a direct consequence of the different tunneling spectra measured when the STM tip sits at a crest or at a valley of the moiré pattern (Figure~\ref{fig:suppl_fig10}(a)). The moiré pattern contrast can be quantified by summing the intensity of the six peaks corresponding to the moiré pattern in the Fourier transforms (FTs) of the conductance maps extracted from $dI/dV(V,\Vec{r})$. This total intensity is plotted as a function of bias voltage in Figure~\ref{fig:figure4}(f) and Figure~\ref{fig:suppl_fig6}(f). The color code is defined in the following way: red when the conductance is larger at the moiré crest, corresponding to the contrast in Figure~\ref{fig:figure4}(a),(d); green when the conductance is larger in the moiré valley, corresponding to the contrast in Figure~\ref{fig:figure4}(b),(e); blue when the conductances are nearly the same at the moiré crest and in the moiré valley, corresponding to the contrast in Figure~\ref{fig:figure4}(c). This bias dependence of the moiré contrast can be directly compared to the numerical difference between the two conductance curves measured at the moiré crest and moiré valley positions in Figure~\ref{fig:suppl_fig10}(a). In  Figure~\ref{fig:suppl_fig10}(c), we plot $dI/dV(V, \text{crest})-dI/dV(V, \text{valley})$, with a positive (negative) result represented by a red (green) dot. This curve is remarkably similar to the one extracted from the Fourier analysis of the conductance maps and confirms the electronic origin of the moiré pattern contrast.

\newpage
\clearpage

\section{Twist angle dependent average charge transfer}
\label{sec:suppl_fig_modeling}

\begin{figure}[htp]
    \centering
    \includegraphics{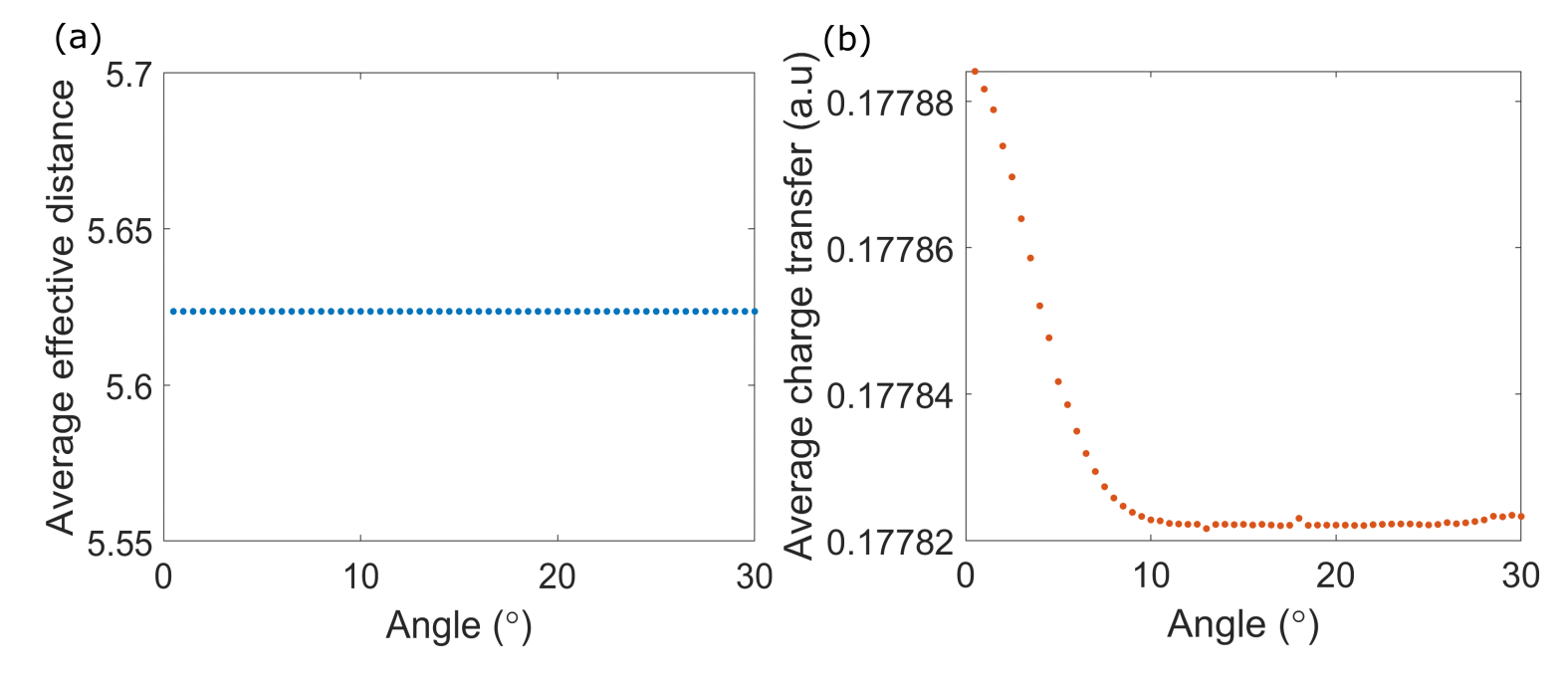}
    \caption{(a) Average of $d_{\text{eff}}(\vec{r}_i)$ and (b) $\overline{\Delta Q}$ in the $1/d_{\text{eff}}$ model as a function of twist angle.}
    \label{fig:suppl_fig_modeling}
\end{figure}

In this model, we consider the charge transfer at a given Au-S nearest neighbor pair to be proportional to $1/d_{\text{eff}}$. We calculate the average charge transfer as:
\begin{equation}
    \overline{\Delta Q}=\frac{1}{N}\sum_{i=1}^{N} \frac{1}{d_{\text{eff}}(\vec{r}_i)}.
\end{equation}
While the average of $d_{\text{eff}}(\vec{r}_i)$ does not change as a function of twist angle (Figure~\ref{fig:suppl_fig_modeling}(a)) we clearly see in Figure~\ref{fig:suppl_fig_modeling}(b) --similarly to the model presented in the main text-- that the average charge transfer is decreasing with increasing twist angle.


 \bibliographystyle{ieeetr}
 \bibliography{biblio.bib}